\documentclass[11pt,a4paper]{article}

\usepackage{graphicx,epsfig,rotating}
\makeatletter
\let\OriginalIncludeGraphics\includegraphics
\renewcommand{\includegraphics}[2][]{%
  \IfFileExists{#2}{%
    \OriginalIncludeGraphics[#1]{#2}%
  }{%
    \begingroup
      \filename@parse{#2}%
      \xdef\safegraphicsfile{\filename@base(1).\filename@ext}%
    \endgroup
    \IfFileExists{\safegraphicsfile}{%
      \OriginalIncludeGraphics[#1]{\safegraphicsfile}%
    }{%
      \OriginalIncludeGraphics[#1]{#2}%
    }%
  }%
}
\makeatother
\usepackage{amsmath}
\usepackage{amssymb}
\usepackage{latexsym}
\DeclareSymbolFontAlphabet{\mathbb}{AMSb}

\begin{document}

\title{ 
Generalized Cartan-Kac Matrices inspired from Calabi-Yau spaces}
\author{E. Torrente-Lujan,\\
GFT, Dept. of Physics, Universidad de Murcia, \\
Spain. email: e.torrente@cern.ch}

\maketitle

\begin{abstract}
The object of this work is the systematical study of a certain 
type of generalized
Cartan matrices associated with the Dynkin diagrams that characterize
Cartan-Lie and affine Kac-Moody algebras.  
These generalized matrices are associated to 
 graphs which arise in the study and classification of 
Calabi-Yau spaces through Toric Geometry.
We focus in the study of what should be considered the 
generalization of the affine exceptional series 
$E_{6,7,8}^{(1)}$ Kac-Moody matrices.
It has been 
 conjectured that these
generalized simply laced graphs and associated link matrices 
may characterize
generalizations of Cartan-Lie and affine Kac-Moody algebras. 
\end{abstract}

\section{Introduction}

Progress in fundamental physics is dependent on the identification of
underlying symmetries such as general coordinate invariance or gauge
invariance. 
The gauge symmetries appearing in the highly successful 
Standard Model (SM) of 
particles and interactions is based on
Cartan-Lie Algebras and their direct products.
There have been valiant efforts to extend the SM within the
framework of Cartan-Lie algebras and with the objective of, for example,
reducing the number of free parameters appearing in the theory. however,
attempts to formulate Grand Unified theories (GUT)
 in which the direct product
of the symmetries of the SM is embedded in some larger simple
Cartan-Lie group have not had the same degree of success as the SM. 
The alternative possibility of unifying the gauge interactions with
gravity in some `Theory of Everything' based on string theory is very
enticing, in particular because this offers novel algebraic structures.

At a very basic level, and without any obvious direct interest for the
content of the SM, 
Cartan-Lie symmetries are closely connected
to the geometry of symmetric homogeneous spaces, which were classified by
Cartan himself. Subsequently, an alternative geometry of non-symmetric
spaces appeared, and their classification was suggested in 1955 by Berger
using holonomy theory~\cite{Berger}. 
There are several infinite series of
spaces with holonomy groups $SO(n)$, $U(n)$, $SU(n)$, $Sp(n)$ and
$Sp(n)\times Sp(1)$, and additionally some exceptional spaces with
holonomy groups $G(2)$, $Spin(7)$, $Spin(16)$.

Beyond the SM, new theories as Superstrings 
 offer new clues how to attack the problem of the
nature of symmetries at a very basic geometric level.  For example, the
compactification of the heterotic string leads to the classification of
states in a representation of the Kac-Moody algebra of the gauge group
$E_8\times E_8$ or $Spin(32)/{ Z}_2$. These structures arose in
compactifications of the heterotic superstring on 6-dimensional Calabi-Yau
spaces, non-symmetric spaces with an $SU(3)$ holonomy group~\cite{CHSW}.
It has been shown~\cite{belavin} that group theory and algebraic
structures play basic roles in the generic two-dimensional conformal field
theories (CFTs) that underlie string theory. The basic ingredients here
are the central extensions of infinite-dimensional Kac-Moody algebras.
There is a clear connection between these algebraic and geometric
generalizations. Affine Kac-Moody algebras are realized as the central
extensions of loop algebras, namely the sets of mappings on a compact
manifold such as $S^1$ that take values on a finite-dimensional Lie
algebra. Superstring theory contains a number of other
infinite-dimensional algebraic symmetries such as the Virasoro algebra
associated with conformal invariance and generalizations of Kac-Moody
algebras themselves, such as hyperbolic and Borcherd algebras.

In connection with Calabi-Yau spaces, (Coxeter-)Dynkin diagrams which are
in one-to-one correspondence with both Cartan-Lie and Kac-Moody algebras
have been revealed through the technique of 
 the crepant
resolution of specific quotient singular structures such as the
Kleinian-Du-Val singularities ${ {\bf C}^2/G}$~\cite{DuVal}, where $G$ is
a discrete subgroup of $SU(2)$.
 Thus,
the rich singularity structure of some examples of non-symmetrical
Calabi-Yau spaces provides another opportunity to uncover
infinite-dimensional affine Kac-Moody symmetries.  The Cartan matrices of
affine Kac-Moody groups are identified with the intersection matrices of
the unions of the complex projective lines resulting from the blow-ups of
the singularities. For example, the crepant resolution of the
${ {\bf C}^2/Z_n}$ singularity gives for rational, i.e., genus-zero, (-2)
curves an intersection matrix that coincides with the A$_{n-1}$ Cartan
matrix.
This is also  the case of $K3 \equiv CY_2$
spaces, where the classification of the degenerations of their 
elliptic 
fibers 
(which can be written in Weierstrass form)
and
their associated singularities leads to a link between $CY_2$ spaces and
the infinite and exceptional series of affine Kac-Moody algebras,
$A_r^{(1)}$, $D_{r}^{(1)}$, $E_6^{(1)}$, $E_7^{(1)}$ and $E_8^{(1)}$ (ADE)~\cite{Kodaira,Ber}.


In the search for new symmetries and from  Calabi-Yau geometry as 
inspirational basis, 
it has been  defined  the so called Berger Matrices \cite{volemi,Vol} 
which generalize Cartan and Cartan-Kac-Moody matrices.
In purely algebraic terms, 
a Berger matrix $B$ is a finite integral matrix characterized by 
the following data:
\begin{eqnarray}
B_{ii}&=&k(i), \ k(i)\in {\mathbb Z}\nonumber\\
B_{ij}& \leq& 0, (i\not = j), \quad B_{ij} \in {\mathbb Z} ,\nonumber\\
B_{ij}=0 &\mapsto & {\mathbb B}_{ji}=0, \nonumber\\
\det \ B &=&0,\quad \det \ B_{\{(i)\}} > 0.\nonumber
\label{eqsberger}
\end{eqnarray}
The last  condition  means  positivity of all 
principal proper minors of the matrix, this, together with the determinant equal zero  corresponds to the  {\it affine condition}. 
A condition  shared by Kac-Moody Cartan matrices.


The concept of Berger matrices is obtained 
  by weakening the conditions on 
the generalized Cartan matrix ${\hat {\mathbb A}}$ appearing in affine 
Kac-Moody algebras. In fact, the only difference
with respect to them is that 
 the usual restriction on the diagonal elements 
($B_{ii}=2$ for CKM matrices) is relaxed. Note that,
more than one type of diagonal entry is 
allowed now: positive integer $2,3,..$ diagonal entries 
can coexist in a given matrix.

{\em ``Non-affine''} Berger Matrices are also defined:
the condition of zero determinant is eliminated. 
These matrices could play in some way   the same role of basic 
simple blocks as finite Lie algebras play for the case 
of affine Kac-Moody algebras.

The study of properties and the 
systematic enumeration of the various 
possibilities concerning the 
large family of possible Berger matrices can be facilitated 
by the introduction for each matrix of its generalized Dynkin diagram
$\Delta (B)$.
A schematic  prescription for the most simple cases could be:
A) For a matrix of dimension $n$, define  $n$ vertexes and  draw them
 as small circles. In case of appearance of vertexes with different 
diagonal entries, some graphical distinction will be performed.
Consider all the element $i,j$ of the matrix in turn.
B) Draw one line from vertex $i$ to vertex $j$ if the corresponding 
element $A_{ij}$ is non zero.

Some essential properties of Cartan and Berger matrices are 
easily deduced from  the 
two well known  Frobenius-Perron Lemmas which we repeat here 
for convenience
\cite{kac}: First lemma.
For a real symmetric matrix $M$ 
with elements $M_{ij,i\not = j}\leq 0$,
If the matrix is positive semi-definite, then
 the smallest eigenvalue of the matrix, eventually a 
zero eigenvalue,
 has multiplicity one and the corresponding 
eigenvector has all positive coordinates.  
Second lemma, in the same case, let $N$ be the 
$(n-1)\times (n-1)$ matrix obtained by deleting the 
$i$ row and column from $M$. Then $N$ is positive definite.

According to these lemma, 
if the  Berger matrix $B$ is affine (has determinant equal to zero) then 
it has one and only one zero eigenvalue and corresponding to it, there is an eigenvector $c$ with 
all positive entries. 
The entries of this vector $c$ are by definition the 
Coxeter labels.
For a Berger matrix $ B$ of dimension $n$, 
the rank is $r=n-1$.
Since $B$ is of rank $r=n-1$, we can find one, and only one, 
non zero vector $\mu$ such that $B\mu=0.$
The numbers, $ a_i$, components of the vector $\mu$, 
are called Coxeter labels.
The sums of the Coxeter labels  $h=\sum \mu_i$ is the 
Coxeter number.
For a symmetric generalized Cartan matrix only this type of 
Coxeter number appear.

The general 
objective of this work is to study, enumerate and
classify all possible matrices of Berger type beyond those already 
known (Cartan and Kac-Moody matrices) and which are particular cases 
of the first ones.
In concrete we study the most basic, yet non trivial, cases, the 
generalization of the exceptional Kac-Moody affine matrices $E_{6,7,8}^{(1)}$.
We choose as a guide the following remark: the matrices of these 
exceptional algebras basically 
consist of blocks containing  standard $A_r$ Cartan 
matrices on them. The corresponding graphs consist of a central 
node from which three legs depart. The number of nodes in each 
leg is strongly restricted. In fact only three possibilities, the 
three exceptional cases, exist.

\section{The Berger generalization of Kac-Moody exceptional matrices}

As a starting example (other examples generalizing $A,D$ series will 
be the object of some other work)
we are interested in Berger matrices which generalize the structure 
$E_{6,7,8}^{(1)}$ matrices. Let us consider Berger matrices 
with the following  block structure:
\begin{eqnarray}
B_{SL}&=&
\begin{pmatrix}  
A_{r_1} &0  &0  &0  &v_1 \\
0 &A_{r_2}  &0  &0  &v_2 \\
0 &0  &A_{r_3} &0  &v_3 \\
0 &0  &0  &A_{r_4}  &v_4 \\
v_1^t &v_2^t  &v_3^t  &v_4^t  &k 
\end{pmatrix}
\label{matrixbsl}
\end{eqnarray}
where $A_{r_i}$  are  Cartan matrices of undefined dimension $r_i$ 
and the $v_i$  are column 
vectors filled with zeroes except for one negative entry, $v_i^t=(0,\dots,0,-1)$. The number $k$ is an arbitrary positive integer.
Obviously these matrices are similar in structure to those matrices 
of the exceptional Kac-Moody affine algebras. The only difference is 
the number of blocks (four instead of three) and the value of $k$.

A general matrix can be obtained considering any number of blocks
$A_r$.
The dimension of the original matrix $B$ is given by
$D=1+\sum_{1,m} r_i$. 
For further reference, let us also define 
another important quantity in algebraic geometry 
applications is the ``squared canonical class'' $K^2$ of the 
matrix $B$. This quantity is defined  as the sum of all 
the elements of the matrix 
$K^2=\sum_{i,j} B_{i,j}.$
For the type of matrices, \ref{matrixbsl}, of interest in this 
work, is important to remark that for Cartan matrices $A_r$, 
$K(A_r)=2.$
Then
$$K^2(B_{SL})=\sum_i K^2(A_{r_i}) - 2m+k=2m-2m+k=k.$$
Finally, let us remark that 
the trace of the matrix $B_{SL}$ is given by 
$tr\  B_{SL}=2 \sum_i r_i+k.$

From this assumed structure it is obviously clear that this 
family of matrices fulfill nearly all the conditions for a Berger 
matrix.
The objective is to find all the sets of values of
 $k,r_1,r_2,r_3,r_4$ which 
make vanish the determinant of the matrix and then check for its 
positive-semidefitiness.

To this purpose, we will compute the determinant of a general matrix 
of the above structure in three different ways.

As a first way, the determinant of a general matrix of the 
type $B_{SL}$ can easily 
be computed by induction  following the Laplace rule along 
 the entries of the column of the $v_i's$.
We have the result (for the $m=4$ case for keeping notation 
simple)
{\small
\begin{eqnarray}
\det B_{SL}&=& 
k\ Q
-\det A_{r_1-1} \det A_{r_2}\det A_{r_3}\det A_{r_4}
-\det A_{r_1} \det A_{r_2-1}\det A_{r_3}\det A_{r_4}\nonumber\\
& &
-\det A_{r_1} \det A_{r_2}\det A_{r_3-1}\det A_{r_4}
-\det A_{r_1} \det A_{r_2}\det A_{r_3}\det A_{r_4-1}
\nonumber\\
&= & 
Q
\left (k
-\frac{\det A_{r_1-1}}{\det A_{r_1}}
-\frac{\det A_{r_2-1}}{\det A_{r_2}}
-\frac{\det A_{r_3-1}}{\det A_{r_3}}
-\frac{\det A_{r_4-1}}{\det A_{r_4}}
 \right )\nonumber\\
Q&= & 
\det A_{r_1}\det A_{r_2}\det A_{r_3}\det A_{r_4}
\label{eqdet}
\end{eqnarray}
}
where the  $A_{r_i-1}$ are matrices of dimension one less than
the original matrix, obtained from them by eliminating the 
last column and  row.
We have $\det A_r = r+1.$ 
 The formula is in this case:
\begin{eqnarray}
\det B_{SL}&=& 
Q
\left ( k
-\frac{r_1}{r_1+1}-\frac{r_2}{r_2+1}
-\frac{r_3}{r_3+1}-\frac{r_4}{r_4+1}
\right )\\
&=&Q
\left ( k-4
+\frac{1}{r_1+1}
+\frac{1}{r_2+1}
+\frac{1}{r_3+1}
+\frac{1}{r_4+1}
\right ), \\
Q&=& 
(r_1+1)(r_2+1)(r_2+1)(r_2+1).
\end{eqnarray}
This is the final formula. From here
we can obtain the finite number of 
combinations of  integers $r_i$ which make 
vanish the determinant. 

We can study the different cases 
according to the value of k. 
\begin{itemize}

\item If $k\geq 4$ then $\det B_{SL}$ is always positive.

\item If $k=3$, th $\det B_{SL}$ can be positive, 
negative  or zero.

The determinant equals  zero if and only if
\begin{eqnarray}
1&=& \frac{1}{r_1+1}
+\frac{1}{r_2+1}
+\frac{1}{r_3+1}
+\frac{1}{r_4+1}.
\end{eqnarray}

If the right side of the expression is less (greater) than one then
the determinant is negative (positive)

\item If $k=2$ the determinant is positive or zero.
The only possibility for the determinant to 
equal zero is that
all $r_i=1$. For $r_i>1$ the determinant is negative.

\item 
Finally, if $k=0,1$ the determinant is always negative.

\end{itemize}

We can easily obtain the behavior in the general case
for a matrix $B_{SL}$ of similar structure as Eq.(\ref{matrixbsl})
but with an arbitrary number $m$ of 
matrices $A_{ri}$ in the diagonal.
It is interesting the particular case  $k=m-1$. In this 
case the condition for the 
vanishing of the determinant is :
\begin{eqnarray}
\sum_{i=1,m}\frac{1}{r_i+1}=1.
\label{detvanish}
\end{eqnarray}
For all the other values of $k\geq m$ the determinant 
is always positive.

In the general case the sign of the determinant is equal 
to the sign of the expression:
\begin{eqnarray}
T&=&k-m+\sum_{i=1,m}\frac{1}{N_i+1}<k-m+\frac{m}{2}=k-\frac{m}{2}.
\end{eqnarray}

For example, for m=5, we have non trivial 
zero conditions for k=4,3.  For non trivial, we mean that 
in that case both expressions
\begin{eqnarray}
\sum_{i=1,m}\frac{1}{r_i+1}&=&1, \quad (k=4), or,\\
\sum_{i=1,m}\frac{1}{r_i+1}&=&2 \quad (k=3) \\
\sum_{i=1,m}\frac{1}{r_i+1}&=&k-m \quad (in general)
\label{e1001}
\end{eqnarray}
have solutions with no all the $r_i$ simultaneously 
equal one.
It is obvious that the combination m=5,k=2 has however no 
zero solutions.

The solution to any of the previous zero conditions will 
be denoted as 
$E_{(r_1,r_2,....r_m)}^{(m-k)}$.

We will see below that Eq.(\ref{e1001}) does not give 
essentially new solutions. Solutions to this equation 
for fixed $k,m$ can be built from the solutions to equations of the same parameter $k$ and different $m_1,m_2$ in an 
iterative way.
For example, below we will obtain the rule
$E^{(m-k_1)}_{(r)}
\triangle 
E^{(m-k_2)}_{(r')}=
E^{(2 m-k_1-k_2)}_{(r')\cup(r)}$.

The integer solutions to equations \ref{e1001}, e.g. 
partitions of unity in terms of ``Egyptian fractions'' are 
well known in arithmetic. These equations can be casted 
in some other way: as partitions of an integer in terms 
of a fixed number of other integers which divide the 
original number. This can be seen as follows.
First, let us define 
$s\equiv l.c.m. (r_1+1,\dots,r_m+1)$ and 
$x_i\equiv s/(r_i+1)$. Then the previous equation
\ref{e1001} is obviously written as
$$\sum_i x_i= (k-m) s.$$ 
Clearly any of the 
summands divides the total: $x_i \mid s$. 
In this formulation, in the most simple case $k-m=1$, 
we are 
interested in numbers $s$ which can written as sum of 
their divisors $x_i$. The number of divisors is fixed, 
$m$, but they can appear repeated. This divisors does not
 need to be prime or relative prime. 
The different solutions will be presented as 
$(x_1,\dots,x_i,\dots)[s]$ in the tables which follows.

\subsection{\bf The Coxeter labels.}

We will re-obtain the same results 
in some slightly different  way. 
The 
advantage is that now we will obtain the values for 
the Coxeter labels and Coxeter number.

According to the Frobenius-Perron lemma, 
if the  Berger-Cartan
  matrix $B$ is affine (has determinant equal to zero) then 
it has one and only one zero eigenvalue and corresponding to it, there is an eigenvector $c$ with 
all positive entries. 
The entries of this vector $c$ are by definition the 
Coxeter labels.

Let us take the previous matrix $B$ and write the vector $c$ in block form corresponding to each of the 
submatrices $A_{ri}$:
$c^t=(c_1,c_2,c_3,c_4,s)$ where the 
subvectors $c_i=(c_{i,1},\dots,c_{i,j},\dots,c_{i,jmax})$,
$j_{max}(i)=r_i$ and $s$ is a, a priori unknown, constant.
We assume that 
$B \ c=0$, this is equivalent to the set of equations:
\begin{eqnarray}
A_{ri} c_i+ s v_{ri} &=& 0, \quad (for\  i=1-4); \\
\sum_i v_i^t \ c_i+k\ s &=& 0.
\label{setcoxeter}
\end{eqnarray}
Due to the special form of the vectors $v_i$, the 
second equation becomes 
$$k\ s=\sum_i c_{i,jmax(i)}.$$

Let us consider now the first set of equations.  
We know already the solution to this kind of equations:
 for a general $A_r$ Cartan matrix
\begin{eqnarray}
A_r \begin{pmatrix}  
1\\ 2\\ \vdots \\ r
    \end{pmatrix}
&=& (r+1)
\begin{pmatrix}
0\\ 0\\ \vdots \\ 1
\end{pmatrix},
\end{eqnarray}
so, we have 
\begin{eqnarray}
c_i^t&=& x_i(1,2,\dots, r_i),\quad 
x_i=\frac{s}{r_i+1}.
\end{eqnarray}

We arrive then to an important conclusion, 
for the 
Coxeter labels to be integers, the constant $s$ is 
multiple of all  the numbers $r_i+1$, where the 
$r_i$ are the 
dimensions of the matrices $A_{ri}$.
Now, we insert this solution in the last equation of the system 
\ref{setcoxeter}. We obtain 
$$c_{i,jmax}=\frac{ s\ r_i}{r_i+1},$$
and then
\begin{eqnarray}
k\ s&=& s \sum_i  \frac{r_i}{r_i+1},\ \quad or, \quad  
k   = m-\sum_i  \frac{1}{r_i+1}\ (if\ s\not = 0).
\end{eqnarray}
 
The constant $s$ has cancelled from the last 
equation and remains arbitrary. The only condition is 
that $s\mid r_i+1$. The minimal choice is given 
by  $$s=lcm (r_1+1,\cdots ,r_i+1,\cdots).$$

An important parameter is the 
Coxeter number $h$ which is defined as the sum of all 
the Coxeter labels. This number is given by
\begin{eqnarray}
h&\equiv&s+\sum_{i,j} c_{i,j}=
 s+\sum_i \frac{s }{r_i+1} \frac{r_i (r_i+1)}{2}\\
 &=& s \left (1+ \frac{1}{2}\sum_i r_i\right ).
\end{eqnarray}
The value of $h$ is minimal for the choice of $s$ above.
We can write the previous formula 
in the simple form including the dimension $D$ of the matrix
\begin{eqnarray}
h&=&\frac{s}{2}\left (1+D\right ).
\end{eqnarray}

\subsection{The determinant and eigenvalues  
of B}

It is  convenient to re-obtain the same results with yet 
another method. As a byproduct of this, the matrix $B$ will 
be diagonalized.
A simple algorithm for this purpose is described 
in Ref.\cite{duchon} (see also Ref.\cite{eisenbud} for 
a description of the method)
to diagonalize 
matrices of the form $B_{SL}(\Delta)$ where 
$\Delta$ is a certain graph (a plumbing tree-like graph)
The diagonalized matrix  has the form 
$D=P^t A(\Delta) P$ with $\det P=1$. Thus $\det D=\det A(\Delta)$. 

The diagonalization of the matrix can be translated into 
a new idea: the diagonalization of the corresponding graph. 
A diagonal graph  is one with isolated nodes, not linked by edges.
Clearly the determinant of the matrix  corresponding to 
the diagonalized graph, where the 
weights has been recalculated in some way,  should be the same 
as the determinant of the initial matrix and graph.

Let us suppose generic trees weighted by arbitrary 
rational numbers $e_i$. 
To diagonalize  a  given such tree $\Delta$, pick 
some vertex (the central node for example) and 
direct all edges toward this vertex. Now 
simplify the graph $\Delta$ by recursively deleting 
edges according to the  procedures presented in 
table 12.3 of Ref.\cite{eisenbud}. 
Eventually we end up with a collection of isolated 
points with new weights $d_i$, this is the diagonalized 
graph $\Delta_D$. Then $D=diag(d_i)$ is 
the desired diagonalization of $A(\Delta)$. 
Thus $\det A(\Delta)=\prod_i d_i $.
In particular, all the weights but the last one are 
guaranteed to be strictly positive numbers. The last 
one is given by:
$$e_f'=e_f-\sum_{i=1,m}\frac{1}{e_i'}.$$
For the case of purely linear subgraphs with 
nodes weighted by $w_j$, each of the $e_i'$ are continued fractions 
of the type 
\begin{eqnarray}
e_i'&=&1-\cfrac{1}{w_1-
\cfrac{1}{w_2-\dotsb}}.
\end{eqnarray}
In the case of interest to us here, obviously 
  we put $e_f=k$ and the quantities $e_i'$ are 
the values of a continued fraction corresponding  
to a linear graph of weight-2 nodes (any of our  $A_r$ sub-graph).
They are easily obtained 
$$e_i'=\frac{r_i}{r_i+1}.$$
Obviously we reobtain the formula given by Eq.\ref{eqdet}. 
In addition we obtain the rest of eigenvalues 
of the matrix and the explicit confirmation that is 
positive-semidefinite.

Note that the continued fraction $e_i'$ correspond to the ratio of two determinants of matrices.  
The quantity $e_f$ is a generalized fraction. It can be interpreted 
as a (generalized) determinant of a set of matrices.

%
%

\subsection{Results: enumeration of cases and conclusions} 

We deal first with the cases with $K^2(B)=k=m-1$. These correspond 
to integer solutions of the equation (\ref{detvanish}).
We consider small values of $m$ ($m\leq 6$).

{ Case m=2, k=m-1=1.}
In this case the condition for 
vanishing  determinant is 
\begin{eqnarray}
1&=&\frac{1}{r_1+1}+\frac{1}{r_2+1},
\end{eqnarray}
the only possibility is the trivial one $r_1=r_2=1$. 
This case 
correspond to the generalized Cartan Matrix:
\begin{eqnarray}
E_{(2,2)}^{(1)}&=&
\begin{pmatrix}
2 & 0 & -1 \\
0 & 2 & -1 \\
-1 & -1 & 1 \\
\end{pmatrix} 
.
\end{eqnarray}

{Case  $m=3, k=m-1=2$.}
In this case the 
condition for vanishing determinant contains 
three ``Egyptian fractions''
\begin{eqnarray}
1&=&\frac{1}{r_1+1}+\frac{1}{r_2+1}+\frac{1}{r_2+1}.
\end{eqnarray}
There are exactly three integer 
solutions to this equation. 
They are
$N=(2,2,2),N=(1,3,3),N=(1,2,5).$
They obviously correspond
to the graphs and Cartan matrices of the 
three exceptional affine Kac-Moody algebras 
$E^{(1)}_{6,7,8}$. In our notation
\begin{eqnarray}
E_{(2,2,2)}^{(1)}&\equiv & E^{(1)}_6, \\
E_{(1,3,3)}^{(1)}&\equiv & E^{(1)}_7, \\
E_{(1,2,5)}^{(1)}&\equiv & E^{(1)}_8.
\end{eqnarray}

Note that the formula for the determinant in this case 
has already be given in the classical book of 
Kac \cite{kac} where the equivalent 
quantities  $T_{r,s,p}$ are computed.

 {Case $m=4,k=m-1=3$.}
In this case the condition for vanishing
 determinant contains 
four Egyptian fractions
\begin{eqnarray}
1&=&\frac{1}{r_1+1}+\frac{1}{r_2+1}+\frac{1}{r_3+1}+\frac{1}{r_4+1},
\end{eqnarray}
there are exactly  fourteen  solutions to this equation, 
They are tabulated in table\ref{t1}.

In the cases $m=5$ and $m=6$ with $k=m-1$ 
 the formulas  for vanishing 
determinant contain summatories of  
five or six Egyptian fractions respectively.
For $m=5$ there are exactly  $147$  solutions to this equation while 
for $m=6$ 
there are exactly  $3240$  solutions to this equation, 
In both case, part of the solutions are tabulated 
in tables (\ref{td5}) and (\ref{td6}) where we 
present those solutions with $D\leq 40$.
Some of them (those with $D\leq 40$). 
For a full list consult Ref.\cite{wwwtorrente}.

For  $m>6$, the condition of vanishing  is similar, 
however in these cases the total number of solutions is not known in 
general. 
However one can easily obtain a large number of solutions by numerical 
methods. It is also obvious that solutions for a given $m$ can 
be obtained from those solutions of an smaller $m$.

\subsection{The cases $k= m-2$}

Let us study some of the solutions obtained for 
$k\not = m-1$ with $m\leq 6$. 

The only non-trivial cases are 
obtained for $m=5,k=3$ and $m=6,k=4$.
For 
$m=5,k=3$ there are exactly 3 solutions which are 
presented in table\ref{t1sds}.
For 
$m=6,k=4$ there are exactly 17 solutions which are 
presented in table\ref{t1wew}.

These solutions are not new in some sense: they can 
be obtained from the solutions of $k=m-1$. 
It is important to remark that apart from them there are 
not any other really new solutions.

From this we can define a binary 
operation on graphs, the 
 ``$\tau$-product'' 
$C=A \stackrel{\otimes}{\tau} B$.
This operation has an important property: if $\det A=0$ and $\det B=0$ 
 then also $\det C=0$.

\begin{figure}
\centering
\begin{tabular}{cc}
\includegraphics[width=0.4\linewidth]{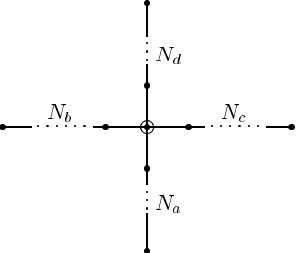} \\[0.8cm]
\includegraphics[width=0.6\linewidth]{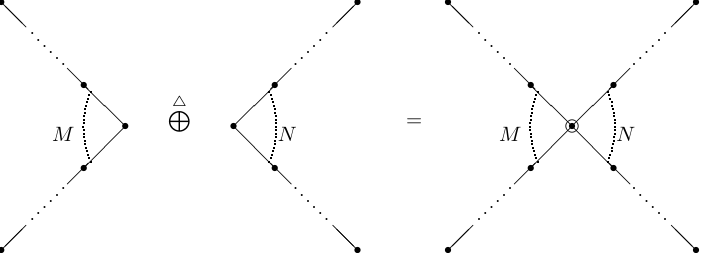}
\end{tabular}
\label{figxxgen}
\caption{(Top) The general graph for simply laced cases.
(Bottom) The symbolic fusion rule. }
\end{figure}

\begin{table}
\centering
{
\begin{tabular}{llccc}
& $E_{(r_a,r_b)}^{(1)}$ & Dim & h  &     \\[0.1cm] \hline
1&$E_{(1,1)}^{(1)}$    & 3    &4    & $(1,1)[2]$    \\
\hline
\end{tabular}
\label{t1}
\caption{ List of $m=2,k=1$}}
\end{table}

\begin{table}
\centering
{\small
\begin{tabular}{llccccc}
& $E_{(r_a,r_b,r_c})$ & Dim & h &  $(...)[s]$ \\[0.1cm]
 \hline
1&$E_{(2,2,2)}^{(1)}\equiv E_{6}^{(1)} $ & 7& 12 & $(1,1,1)[3]$   \\[0.1cm]
2&$E_{(1,3,3)}^{(1)}\equiv E_{7}^{(1)}$ & 8& 18 & $(1,1,2)[4]$  \\[0.1cm]
3&$E_{(1,2,5)}^{(1)}\equiv E_{8}^{(1)}$ & 9& 30 & $(1,2,3)[6]$  \\[0.1cm]

\hline
\end{tabular}
\label{t1b}
\caption{ List of of cases with 
 $m=3,k=2$. They correspond to the 
standard Kac-Moody exceptional algebras 
(cf. with $T_{qrs}$
 in table 2.3 in Ref.\protect\cite{kac}.) }}
\end{table}

\begin{table}
\centering
{\small
\begin{tabular}{llccc}
& $E_{(r_a,r_b,r_c,r_d)}^{(1)}$ & Dim & h & $(...)[s]$ \\[0.1cm] \hline
1&$E_{(3,3,3,3)}^{(1)}$ & 13&28 & $(1,1,1,1)[4]$   \\
2&$E_{(2,3,3,5)}^{(1)}$ & 14&90 &$(2,3,3,4)[12]$  \\
3&$E_{(1,5,5,5)}^{(1)}$ & 17&54 & $(1,1,1,3)[6]$ \\
4&$E_{(2,2,5,5)}^{(1)}$ & 15&48 &$(1,1,2,2)[6]$     \\
5&$E_{(1,3,7,7)}^{(1)}$ & 19&80 & $(1,1,2,4)[8]$  \\
6&$E_{(1,4,4,9)}^{(1)}$ & 19&100& $(1,2,2,5)[10]$\\
7&$E_{(1,3,5,11)}^{(1)}$& 21&132& $(1,2,3,6)[12]$ \\
8&$E_{(2,2,3,11)}^{(1)}$& 19&120& $(1,3,4,4)[12]$ \\
9&$E_{(1,3,4,19)}^{(1)}$& 28&290& $(1,4,5,10)[20]$  \\
10&$E_{(1,2,11,11)}^{(1)}$&26 &162&$(1,1,4,6)[12]$   \\
11&$E_{(1,2,8,17)}^{(1)}$ &29 &270&$(1,2,6,9)[18]$   \\
12&$E_{(1,2,7,23)}^{(1)}$ & 34&420&$(1,3,8,12)[24]$\\
13&$E_{(1,2,9,14)}^{(1)}$&27&420&$(2,3,10,15)[30]$ \\
14&$E_{(1,2,6,41)}^{(1)}$&51&1092 & $(1,6,14,21)[42]$\\
\hline
\end{tabular}
}
\label{t1c}
\caption{ List of  all the 14 solutions for the case
 $m=4,k=3$. The integers $(r_a,r_b,r_c,r_d)$ define both, 
the number of nodes in each of the four legs of the graph
 and the dimension of the each of the block matrices 
$A_r$.
The dimension of the matrix ($D=rank+1=1+\sum r_i$).
 The Coxeter number $h=s/2 (1+D)$.
In the last column appears the form of the solution
$\sum_i x_i =(k-m) s$.}
\end{table}

\begin{table}
\centering
{\small
\begin{tabular}{llccc}
& $E_{(\dots,N_i,\dots )}^{(2)}$ & Dim & h & \\[0.15cm] 
 \hline
1&$E_{(1,1,2,2,2)}^{(2)}\equiv E_{(1,1)}^{(1)}\ \triangle\  E_{(2,2,2)}^{(1)}  $ & 9& 30 & $(2,2,2,3,3)[6]$   \\[0.15cm]
2&$E_{(1,1,1,3,3)}^{(2)}\equiv E_{(1,1)}^{(1)}\ \triangle\ E_{(1,3,3)}^{(1)} $ & 10& 20 & $(1,1,4,4,4)[4]$  \\[0.15cm]
3&$E_{(1,1,1,2,5)}^{(2)}\equiv E_{(1,1)}^{(1)}\ \triangle\ E_{(1,2,5)}^{(1)} $ 
& 11& 36 &$(1,2,3,3,3)[6]$ \\[0.1cm]
\hline
\end{tabular}
}
\label{t1sds}
\caption{ List of  $m=5,k=m-2=3$.}
\end{table}

\begin{table}
\centering
{\scriptsize
\begin{tabular}{lr}
\begin{tabular}{lllrr}
& $E_{(\dots,r_i,\dots) }^{(1)}$ & Dim & h & $(...)[s]$ \\[0.1cm] \hline
1&$E_{(4,4,4,4,4)}^{(1)}$   &21& 55&  $(1,1,1,1,1)[5 ]$\\ 
2&$E_{(3,3,5,5,5)}^{(1)}$   &22&138 &$(2,2,2,3,3)[12 ]$\\ 
3&$E_{(2,5,5,5,5)}^{(1)}$   &23&72 &$(1,1,1,1,2)[6 ]$\\ 
4&$E_{(3,3,3,7,7)}^{(1)}$   &24&100 & $(1,1,2,2,2)[8 ]$\\ 
5&$E_{(3,3,4,4,9)}^{(1)}$   &24&250 & $(2,4,4,5,5)[20 ]$\\ 
6&$E_{(2,3,5,7,7)}^{(1)}$   &25&312 & $(3,3,4,6,8)[24 ]$\\ 
7&$E_{(2,4,4,5,9)}^{(1)}$   &25&390 & $(3,5,6,6,10)[30 ]$\\ 
8&$E_{(3,3,3,5,11)}^{(1)}$  &26&162 & $(1,2,3,3,3)[12 ]$\\ 
9&$E_{(2,3,5,5,11)}^{(1)}$  &27&168 & $(1,2,2,3,4)[12 ]$\\ 
10&$E_{(2,2,8,8,8)}^{(1)}$  &29&135 & $(1,1,1,3,3)[9 ]$\\ 
11&$E_{(2,4,4,4,14)}^{(1)}$ &29&225 & $(1,3,3,3,5)[15 ]$\\ 
12&$E_{(1,7,7,7,7)}^{(1)}$  &30&124 & $(1,1,1,1,4)[8 ]$\\ 
13&$E_{(2,2,7,7,11)}^{(1)}$ &30&372 & $(2,3,3,8,8)[24 ]$\\ 
14&$E_{(1,5,8,8,8)}^{(1)}$  &31&288 & $(2,2,2,3,9)[18 ]$\\ 
15&$E_{(2,3,3,11,11)}^{(1)}$&31&192 & $(1,1,3,3,4)[12 ]$\\ 
16&$E_{(1,5,7,7,11)}^{(1)}$ &32&396 & $(2,3,3,4,12)[24 ]$\\ 
17&$E_{(2,2,5,11,11)}^{(1)}$&32&198 & $(1,1,2,4,4)[12 ]$\\ 
18&$E_{(2,3,3,9,14)}^{(1)}$ &32&495 & $(4,6,15,15,20)[60 ]$\\ 
19&$E_{(1,4,9,9,9)}^{(1)}$  &33&170 & $(1,1,1,2,5)[10 ]$\\ 
\hline
\end{tabular}
&
\begin{tabular}{llllr}
& $E_{(\dots,r_i,\dots) }^{(1)}$ & Dim & h & $(...)[s]$ \\[0.1cm] \hline
20&$E_{(1,6,6,6,13)}^{(1)}$ &33&238 & $(1,2,2,2,7)[14 ]$\\ 
21&$E_{(2,2,5,9,14)}^{(1)}$ &33&510 & $(2,3,5,10,10)[30 ]$\\ 
22&$E_{(3,3,3,4,19)}^{(1)}$ &33&340 & $(1,4,5,5,5)[20 ]$\\ 
23&$E_{(1,5,5,11,11)}^{(1)}$&34&210 & $(1,1,2,2,6)[12 ]$\\ 
24&$E_{(2,3,3,8,17)}^{(1)}$ &34&630 & $(2,4,9,9,12)[36 ]$\\ 
25&$E_{(2,3,4,5,19)}^{(1)}$ &34&1050 & $(3,10,12,15,20)[60 ]$\\ 
26&$E_{(1,5,5,9,14)}^{(1)}$ &35&540 & $(2,3,5,5,15)[30 ]$\\ 
27&$E_{(2,2,5,8,17)}^{(1)}$ &35&324 & $(1,2,3,6,6)[18 ]$\\ 
28&$E_{(1,5,5,8,17)}^{(1)}$ &37&342 & $(1,2,3,3,9)[18 ]$\\ 
29&$E_{(2,2,4,14,14)}^{(1)}$&37&285 & $(1,1,3,5,5)[15 ]$\\ 
30&$E_{(2,2,6,6,20)}^{(1)}$ &37&399 & $(1,3,3,7,7)[21 ]$\\ 
31&$E_{(1,3,11,11,11)}^{(1)}$&38&234 & $(1,1,1,3,6)[12 ]$\\ 
32&$E_{(1,3,9,11,14)}^{(1)}$&39&1200 & $(4,5,6,15,30)[60 ]$\\ 
33&$E_{(1,4,5,14,14)}^{(1)}$&39&600 & $(2,2,5,6,15)[30 ]$\\ 
34&$E_{(1,4,7,7,19)}^{(1)}$ &39&800 & $(2,5,5,8,20)[40 ]$\\ 
35&$E_{(1,5,6,6,20)}^{(1)}$ &39&840 & $(2,6,6,7,21)[42 ]$\\ 
36&$E_{(2,2,4,11,19)}^{(1)}$&39&1200 & $(3,5,12,20,20)[60 ]$\\ 
37&$E_{(2,3,3,7,23)}^{(1)}$ &39&480 & $(1,3,6,6,8)[24 ]$\\ 
38&$E_{(2,2,5,7,23)}^{(1)}$ &40&492 & $(1,3,4,8,8)[24 ]$ \\
\hline
\end{tabular}
\end{tabular}

\label{td5}
\caption{ List of  the $dim\leq 40$ cases of the total of  147 
cases of dimension five $m=5,k=4$.
}
}
\end{table}

\begin{table}{
\centering
{\scriptsize
\begin{tabular}{lr}
\begin{tabular}{lllrr}
& $E_{(\dots,r_i,\dots)}^{(1)}$ & Dim & h & $(..)[s]$\\[0.1cm] \hline
1&$E_{(5,5,5,5,5,5)}^{(1)}$ &31&96 & $(1,1,1,1,1,1)[6 ]$\\ 
2&$E_{(3,5,5,5,7,7)}^{(1)}$ &33& 408& $(3,3,4,4,4,6)[24 ]$\\ 
3&$E_{(4,4,5,5,5,9)}^{(1)}$ &33&510 & $(3,5,5,5,6,6)[30 ]$\\ 
4&$E_{(3,3,7,7,7,7)}^{(1)}$ &35&144 & $(1,1,1,1,2,2)[8 ]$\\ 
5&$E_{(3,4,4,7,7,9)}^{(1)}$ &35&720 & $(4,5,5,8,8,10)[40 ]$\\ 
6&$E_{(3,5,5,5,5,11)}^{(1)}$&35&216 & $(1,2,2,2,2,3)[12 ]$\\ 
7&$E_{(4,4,4,4,9,9)}^{(1)}$ &35&180 & $(1,1,2,2,2,2)[10 ]$\\ 
8&$E_{(2,5,7,7,7,7)}^{(1)}$ &36&444 & $(3,3,3,3,4,8)[24 ]$\\ 
9&$E_{(3,3,5,8,8,8)}^{(1)}$ &36&666 & $(4,4,4,6,9,9)[36 ]$\\ 
10&$E_{(2,5,5,8,8,8)}^{(1)}$ &37&342 & $(2,2,2,3,3,6)[18 ]$\\ 
11&$E_{(3,3,5,7,7,11)}^{(1)}$&37&456 & $(2,3,3,4,6,6)[24 ]$\\ 
\hline
\end{tabular}
&
\begin{tabular}{lllrr}
& $E_{(\dots,r_i,\dots)}^{(1)}$ & Dim & h & $(..)[s]$\\[0.1cm] \hline
12&$E_{(3,4,4,5,9,11)}^{(1)}$&37&1140 & $(5,6,10,12,12,15)[60 ]$\\ 
13&$E_{(4,4,4,5,5,14)}^{(1)}$&37&570 & $(2,5,5,6,6,6)[30 ]$\\ 
14&$E_{(2,5,5,7,7,11)}^{(1)}$&38&468 & $(2,3,3,4,4,8)[24 ]$\\ 
15&$E_{(3,3,4,9,9,9)}^{(1)}$ &38&390 & $(2,2,2,4,5,5)[20 ]$\\ 
16&$E_{(3,3,6,6,6,13)}^{(1)}$&38&546 & $(2,4,4,4,7,7)[28 ]$\\ 
17&$E_{(2,4,5,9,9,9)}^{(1)}$ &39&600 & $(3,3,3,5,6,10)[30 ]$\\ 
18&$E_{(2,5,6,6,6,13)}^{(1)}$&39&840 & $(3,6,6,6,7,14)[42 ]$\\ 
19&$E_{(3,3,5,5,11,11)}^{(1)}$&39&240 & $(1,1,2,2,3,3)[12 ]$\\ 
20&$E_{(2,5,5,5,11,11)}^{(1)}$&40&246 & $(1,1,2,2,2,4)[12 ]$\\ 
21&$E_{(3,3,5,5,9,14)}^{(1)}$ &40&1230 & $(4,6,10,10,15,15)[60 ]$\\
\hline
\end{tabular}

\end{tabular}

}
\label{td6}
\caption{ $m=6,k=5$. List of  all the $dim\leq 40$ cases of a total of 3120 cases.
}}
\end{table}

\begin{table}
\centering
{\scriptsize
\begin{tabular}{lllrr}
& $E_{(\dots,N_i,\dots) }^{(1)}$ & Dim & h & $(...)[s]$ 
\\[0.1cm] \hline
1 &$E_{(1,1,3,3,3)}^{(2)}    \equiv E_{(1,1)}^{(1)} \triangle  E_{(3,3,3,3)}^{(1)}$   & 15 & 32  & $(1,1,1,1,2,2)[4]$   \\[0.1cm]
  &$\phantom{E_{(1,1,3,3,3)}^{(2)}}    \equiv E_{(1,3,3)}^{(1)} \triangle  E_{(1,3,3)}^{(1)}$   &  &   &   \\[0.1cm]
2 &$E_{(1,1,2,3,3,5)}^{(2)}  \equiv E_{(1,1)}^{(1)} \triangle  E_{(2,3,3,5)}^{(1)}$   & 16 & 102 &$(2,3,3,4,6,6)[12]$  \\[0.1cm]
  &$\phantom{E_{(1,1,2,3,3,5)}^{(2)}}  \equiv E_{(1,2,5)}^{(1)} \triangle  E_{(1,3,3)}^{(1)}$   &  & &  \\
3 &$E_{(1,1,1,5,5,5)}^{(2)}  \equiv E_{(1,1)}^{(1)} \triangle  E_{(1,5,5,5)}^{(1)}$   & 19 & 60  & $(1,1,1,3,3,3)[6]$ \\
4 &$E_{(1,1,2,2,5,5)}^{(2)}  \equiv E_{(1,1)}^{(1)} \triangle  E_{(2,2,5,5)}^{(1)}$   & 17 & 54  &$(1,1,2,2,3,3)[6]$     \\
  &$\phantom{E_{(1,1,2,2,5,5)}^{(2)}}  \equiv E_{(1,2,5)}^{(1)} \triangle  E_{(1,2,5)}^{(1)}$   &  &   &    \\
5 &$E_{(1,1,1,3,7,7)}^{(2)}  \equiv E_{(1,1)}^{(1)} \triangle  E_{(1,3,7,7)}^{(1)}$   & 21 & 88  & $(1,1,2,4,4,4)[8]$  \\
6 &$E_{(1,1,1,4,4,9)}^{(2)}  \equiv E_{(1,1)}^{(1)} \triangle  E_{(1,4,4,9)}^{(1)}$   & 21 & 110 & $(1,2,2,5,5,5)[10]$\\
7 &$E_{(1,1,1,3,5,1)}^{(2)}  \equiv E_{(1,1)}^{(1)} \triangle  E_{(1,3,5,11)}^{(1)}$  & 23 & 144 & $(1,2,3,6,6,6)[12]$ \\
8 &$E_{(1,1,2,2,3,1)}^{(2)}  \equiv E_{(1,1)}^{(1)} \triangle  E_{(2,2,3,11)}^{(1)}$  & 21 & 132 & $(1,3,4,4,6,6)[12]$ \\
9 &$E_{(1,1,1,3,4,19)}^{(2)} \equiv E_{(1,1)}^{(1)} \triangle  E_{(1,3,4,19)}^{(1)}$  & 30 & 310 & $(1,4,5,10,10,10)[20]$  \\
10&$E_{(1,1,1,2,11,11)}^{(2)}\equiv E_{(1,1)}^{(1)} \triangle  E_{(1,2,11,11)}^{(1)}$ & 28 & 174 &$(1,1,4,6,6,6)[12]$   \\
11&$E_{(1,1,1,2,8,17)}^{(2)} \equiv E_{(1,1)}^{(1)} \triangle  E_{(1,2,8,17)}^{(1)}$  & 31 & 288 &$(1,2,6,9,9,9)[18]$   \\
12&$E_{(1,1,1,2,7,23)}^{(2)} \equiv E_{(1,1)}^{(1)} \triangle  E_{(1,2,7,23)}^{(1)}$  & 36 & 444 &$(1,3,8,12,12,12)[24]$\\
13&$E_{(1,1,1,2,9,14)}^{(2)} \equiv E_{(1,1)}^{(1)} \triangle  E_{(1,2,9,14)}^{(1)}$  & 29 & 450 &$(2,3,10,15,15,15)[30]$ \\
14&$E_{(1,1,1,2,6,41)}^{(2)} \equiv E_{(1,1)}^{(1)} \triangle  E_{(1,2,6,41)}^{(1)}$  & 53 & 1134& $(1,6,14,21,21,21)[42]$\\
15&$E_{(2,2,2,2,2,2)}^{(2)} \equiv E_{(2,2,2)}^{(1)} \triangle  E_{(2,2,2)}^{(1)}$  & 13 & 21& $(1,1,1,1,1,1)[3]$\\
16&$E_{(1,2,5,2,2,2)}^{(2)} \equiv E_{(1,2,5)}^{(1)} \triangle  E_{(2,2,2)}^{(1)}$  & 14 & 42& $(2,2,3,3,3,6)[6]$\\
17&$E_{(1,3,3,2,2,2)}^{(2)} \equiv E_{(1,3,3)}^{(1)} \triangle  E_{(2,2,2)}^{(1)}$  
& 15 & 96& $(2,4,4,4,4,6)[12]$\\
\hline
\end{tabular}
}
\label{t1wew}
\caption{the $m=6,k=m-2=4$. List of  all the distintct 17 cases.}
\end{table}

\begin{figure}[htb]
\centering
\begin{tabular}{lcrr}
\includegraphics[width=.2\linewidth]{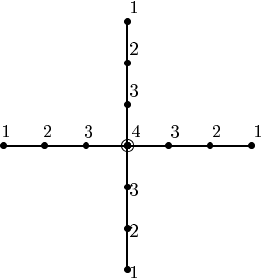} &
\includegraphics[width=.2\linewidth]{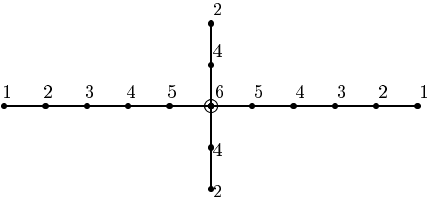} &
\includegraphics[width=.2\linewidth]{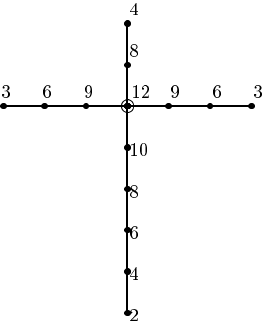}&
\includegraphics[width=.2\linewidth]{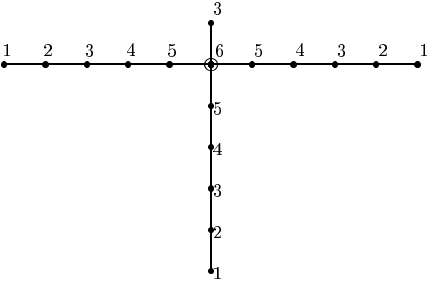} \\
\includegraphics[width=.2\linewidth]{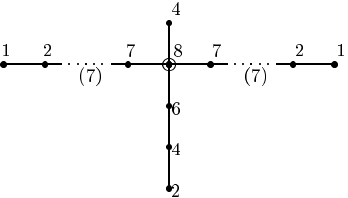} &
\includegraphics[width=.2\linewidth]{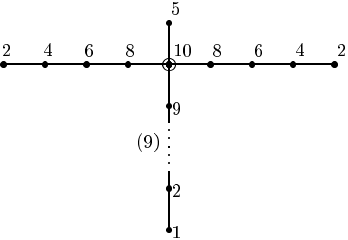} &
\includegraphics[width=.2\linewidth]{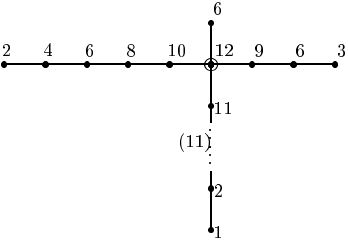} &
\includegraphics[width=.2\linewidth]{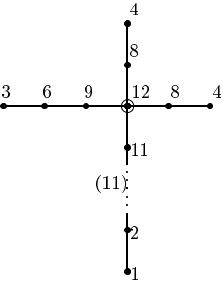} \\
\includegraphics[width=.2\linewidth]{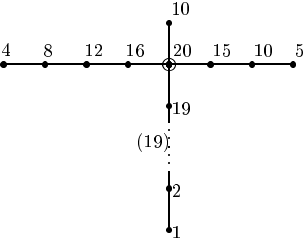} &
\includegraphics[width=.2\linewidth]{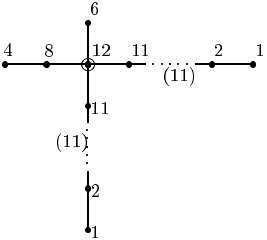} &
\includegraphics[width=.2\linewidth]{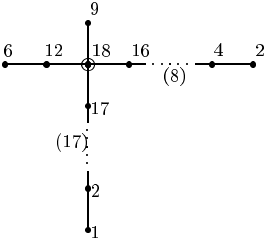} &
\includegraphics[width=.2\linewidth]{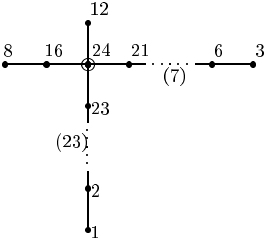}\\
\includegraphics[width=.2\linewidth]{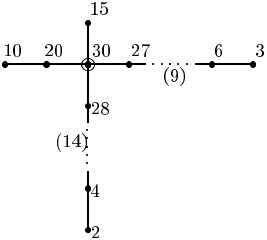} &
\includegraphics[width=.2\linewidth]{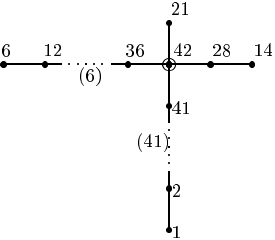} 
\end{tabular}
\label{figalla14}
\caption{all 14 figures}
\end{figure}

In conclusion,
the interest to look for new algebras  beyond Lie algebras started 
from   the $SU(2)$- conformal theories 
(see for example \cite{CIZ,FZ}).
One can think that  geometrical concepts, in particular algebraic geometry,
 could be a  natural and more promising way to  do this.
This marriage of algebra and geometry has been useful in both ways.

It is very  well known, by the Serre theorem,
  that Dynkin diagrams defines  one-to-one   Cartan matrices and these ones Lie or Kac-Moody 
algebras.
In this work, we have generalized    
 some of the properties of Cartan matrices 
for Cartan-Lie and Kac-Moody algebras  into a new class of affine, and non-affine Berger matrices.
We arrive then to the obvious conclusion that 
any algebraic structure emerging from this 
can not be a CLA or KMA  algebra. 
The main difference of these matrices with respect 
previous definitions being in the values that 
diagonal elements of the matrices can take. 

In this work we have enumerated in a systematic way all the 
possibilities of a special class of Berger matrices which 
includes as a subclass the ordinary matrices 
of Kac-Moody affine exceptional 
algebras.
{\small
\vspace{0.6cm}
{\bf Acknowledgments}.
We  acknowledge the  financial  support of 
 the  Spanish CYCIT  funding agency (Ministerio de Ciencia y Tecnologia)
  and the CERN 
Theoretical Division. 
We also  acknowledge the kind hospitality of the 
Dept. of Theoretical Physics, C-XI of the U. Autonoma de Madrid.

}

\end{document}